\documentclass[final,authoryear,5p,times,twocolumn]{elsarticle}

\usepackage{graphicx,fixltx2e}
\usepackage{lineno}
\usepackage{amssymb,amsmath,textcomp}

\journal{Ocean Modelling}

\begin{document}

\begin{frontmatter}

\title{Oceanic three-dimensional Lagrangian Coherent
Structures:\\
A study of a mesoscale eddy in the Benguela ocean region.}

\author{Jo\~ao H. Bettencourt\corref{cor1}\fnref{phone}}
\ead{joaob@ifisc.uib-csic.es}
\cortext[cor1]{Corresponding author}
\fntext[phone]{Phone: +34 971259905}

\author{Crist\'obal L\'{o}pez}
\author{Emilio Hern\'{a}ndez-Garc\'{\i}a}

\address{IFISC (CSIC-UIB), Instituto de F\'{\i}sica Interdisciplinar y Sistemas Complejos  \\
              Campus Universitat de les Illes Balears\\
              E-07122 Palma de Mallorca, Spain}

\begin{abstract}

We study three dimensional oceanic Lagrangian Coherent
Structures (LCSs) in the Benguela region, as obtained from an
output of the ROMS model.
To do that
we first compute
Finite-Size Lyapunov exponent (FSLE) fields in the region
volume, characterizing mesoscale stirring and mixing.
Average FSLE values show a general decreasing trend with depth,
but there is a local maximum at about 100 m depth. LCSs are
extracted as ridges of the calculated FSLE fields. They present
a ``curtain-like" geometry in which the strongest attracting
and repelling structures appear as quasivertical surfaces. LCSs
around a particular cyclonic eddy, pinched off from the
upwelling front are also calculated. The LCSs
are confirmed to
provide pathways and barriers to transport in and out of the
eddy.

\end{abstract}


\end{frontmatter}

\linenumbers
\setlength\linenumbersep{7pt}

\section{Introduction.}
\label{intro}

Mixing and transport processes are fundamental to determine the
physical, chemical and biological properties of the oceans.
From plankton dynamics to the evolution of pollutant spills,
there is a wide range of practical issues that benefit from a
correct understanding and modeling of these processes. Although
mixing and transport in the oceans occur in a wide range of
scales, mesoscale and sub-mesoscale variability are known to
play a very important role \citep{Mahadevan2008,Klein2009}.

Mesoscale eddies are especially important in this aspect
because of their long life in oceanic flows, and  their
stirring and mixing properties. In the southern Benguela, for
instance, cyclonic eddies shed from the Agulhas current can
transport and exchange warm waters from the Indian Ocean to the
South Atlantic \citep{Byrne1995,Lehahn2011}.
 Moreover, mesoscale
eddies have been shown to drive important biogeochemical
processes in the ocean such as the vertical flux of nutrients
into the euphotic zone \citep{McGillicuddy1998,Oschlies1998}.
Another effect of
eddy activity seems to be the intensification
of mesoscale and sub-mesoscale variability due to the
filamentation process where strong tracer gradients are created
by the stretching of tracers in the shear- and strain-dominated
regions in between eddy cores \citep{Elhmaidi1993}.
Studies of the vertical structure of such eddies in the
Benguela region (e. g.
\citet{Doglioli2007} and \citet{Rubio2009288}) have shown that
they can extended to one thousand meters deep waters.

In the last decades new developments in the description and
modelling of oceanic mixing and transport from a Lagrangian
viewpoint have emerged \citep{Mariano2002,Lacasce2008}.
These Lagrangian approaches have become more and more frequent
due to
the increased availability of detailed knowledge of the
velocity field from Lagrangian drifters, satellite measurements
and  computer models. In particular, the very relevant concept
of Lagrangian Coherent Structure (LCS)
\citep{Haller2000,Haller2000b} is becoming crucial for the
analysis of transport in flows. LCSs are structures that
separate regions of the flow with different dynamical behavior.
They give a general geometric view of the dynamics, acting as a
(time-dependent) roadmap for the flow. They are templates
serving as proxies to, for instance, barriers and avenues to
transport or eddy boundaries
\citep{Boffetta2001,Haller2000b,Haller2002,dOvidio2004,dOvidio2009,
Mancho2006b}.

The relevance of the three-dimensional structure of LCSs begins
to be unveiled in atmospheric contexts
\citep{DuToit2010,Tang2011,Tallapragada2011}. In the case of
oceanic flows, however, the identification of the LCSs and the
study of their role on biogeochemical tracers transport has
been mostly restricted to the marine surface
\citep{dOvidio2004,Waugh2006, dOvidio2009,BeronVera2008}. This
is mainly due to two reasons: a) tracer vertical displacement
is usually very small with respect to the horizontal one; and
b) satellite data of any quantity (temperature, chlorophyll,
altimetry for velocity, etc..) are only available from the
observation of the ocean surface.

Oceanic flows can be considered mainly two-dimensional, because
there is a great disparity between the horizontal and vertical
length scales, and they are strongly stratified due to the
Earth's rotation. There are, however, areas in the ocean where
vertical motions are fundamental. Firstly there are the
so-called upwelling regions, which are the most biologically
active marine zones in the world \citep{Rossi2008,Pauly1995}.
The reason is that due to an Ekmann pumping mechanism close to
the coast, there is a surface uprising of deep cold waters rich
in nutrients, inducing a high proliferation of plankton
concentration. Typically, vertical velocities in upwelling
regions are much larger than in open ocean, but still one order
of magnitude smaller than horizontal velocities. Another
example where there are significant vertical processes are
mesoscale eddies producing submesoscale structures
(frontogenesis), which are responsible for strong ageostrophic
vertical process, in addition to the vertical exchange thought
to occur at the eddy interior \citep{Klein2009}. Thus, the
identification of the three-dimensional (3d) LCSs in these
areas is crucial, as well as understanding their correlations
with biological activity. Another reason to include the third
dimension in LCS studies is to investigate the vertical
variation in their properties.

The  main objective of this
paper is the characterization of 3d LCSs, extracted in an
upwelling region, the Benguela area in the Southern Atlantic
Ocean. For this goal we use Finite-Size Lyapunov Exponents
(FSLEs). FSLEs \citep{Aurell1997,Artale1997} measure the
separation rate of fluid particles between two given distance
thresholds. LCSs are computed as the ridges of the FSLE field
\citep{dOvidio2004,Molcard2006,Haza2008,dOvidio2009,Poje2010,Haza2010}.
The rigorous definition of LCS as ridges of a Lagrangian
stretching measure was given for the Finite-Time Lyapunov
Exponents (FTLE) in \citet{Shadden2005} and \citet{Lekien2007},
which are closely related to FSLEs. More recently, hyperbolic LCS
have been defined independently of such stretching measures by
\citet{Haller2011}.
Following many previous studies \citep{dOvidio2004,Molcard2006,
dOvidio2009,Branicki2009} we adopt the mathematical results
for Finite-Time Lyapunov Exponents (FTLE) to FSLE, assuming
them to be valid. In particular, we assume that LCS are identified
with ridges \citep{Haller2001}, i.e., the local extrema of
the FTLE field, and also we expect, in accordance to the
results in \citet{Shadden2005} and \citet{Lekien2007} for FTLEs, that the
material flux through these LCS is small and that they are
transported by the flow as quasi-material surfaces.

To confirm that our identification of LCSs
with ridges of the FSLE field, we perform (in Sect. III) direct
particle trajectory integrations that show that the computed
LCS really organize the tracer flow. In our work, we will
emphasize the numerical methodology since up to now FSLEs have
only been computed for the marine surface (an exception is
\citet{Ozgokmen2011}). We then focus on a particular eddy very
prominent in the area at the chosen temporal window and study
the stirring and mixing on it's vicinity. Some previous results
for Lagrangian eddies were obtained by \citet{Branicki2010} and
\citet{Branicki2011}, applying the methodology of lobe dynamics
and the turnstile mechanism to eddies pinched off from the Loop
Current. In this paper we focus on FSLE fields and the
associated particle trajectories to study transport in and out
of the chosen mesoscale eddy. Since this is a first attempt to
study 3d oceanic LCS, more general results (on Benguela and
other upwelling regions) are left for future work.

To circumvent the lack of appropriate observational data in the
vertical direction, we use velocity fields from a numerical
simulation.
They are high resolution simulations from the ROMS model (see section
\ref{sec:data} below) thus
appropriate to study regional-medium scale basins.

The paper is organized as follows:
In section II we describe the data and methods. In section III
we present our results. Section IV contains a discussion of the
results and Section V summarizes our conclusions.

\section{Data and Methods.}
\label{sec:data}

\subsection{Velocity data set.}
\label{subsec:velocity}

The Benguela ocean region is situated off the west coast of
southern Africa. It is characterized by a vigorous coastal
upwelling regime forced by equatorward winds, a substantial
mesoscale activity of the upwelling front in the form of eddies
and filaments, and also by the northward drift of Agulhas
eddies.

The velocity data set  comes from a regional ocean model
simulation of the Benguela Region \citep{LeVu2011}. ROMS
\citep{Shchepetkin2003,Shchepetkin2005} is a split-explicit
free-surface, topography following model. It solves the incompressible
primitive equations
using the Boussinesq and hydrostatic approximations. Potential
temperature and salinity transport are included by coupling
advection/diffusion schemes for these variables. The model was
forced with climatological data. The data set area extends from
12\textdegree S to 35\textdegree S and from 4\textdegree E to
19\textdegree E (see Fig. 1). The velocity field $\mathbf{u}=(u,v,w)$
consists of two years of daily averaged zonal ($u$),
meridional ($v$), and vertical velocity ($w$) components,
stored in a three-dimensional grid with an horizontal
resolution of $1/12$ degrees $\sim 8$ km, and $32$ vertical
terrain-following levels
using a stretched vertical  coordinate where the
layer thickness varies, increasing from the surface
to the ocean interior.
Since the ROMS model considers the hydrostatic approximation
it is important to note that
\citet{Amala2006222}, when
comparing results from non-hydrostatic and hydrostatic
versions of the same model of vertical motions
at submesoscale fronts, found that while instantaneous
vertical velocities structures differ, the averaged
vertical flux is similar in both hydrostatic and non-hydrostatic
simulations.

\begin{figure}
  \includegraphics[width=\columnwidth]{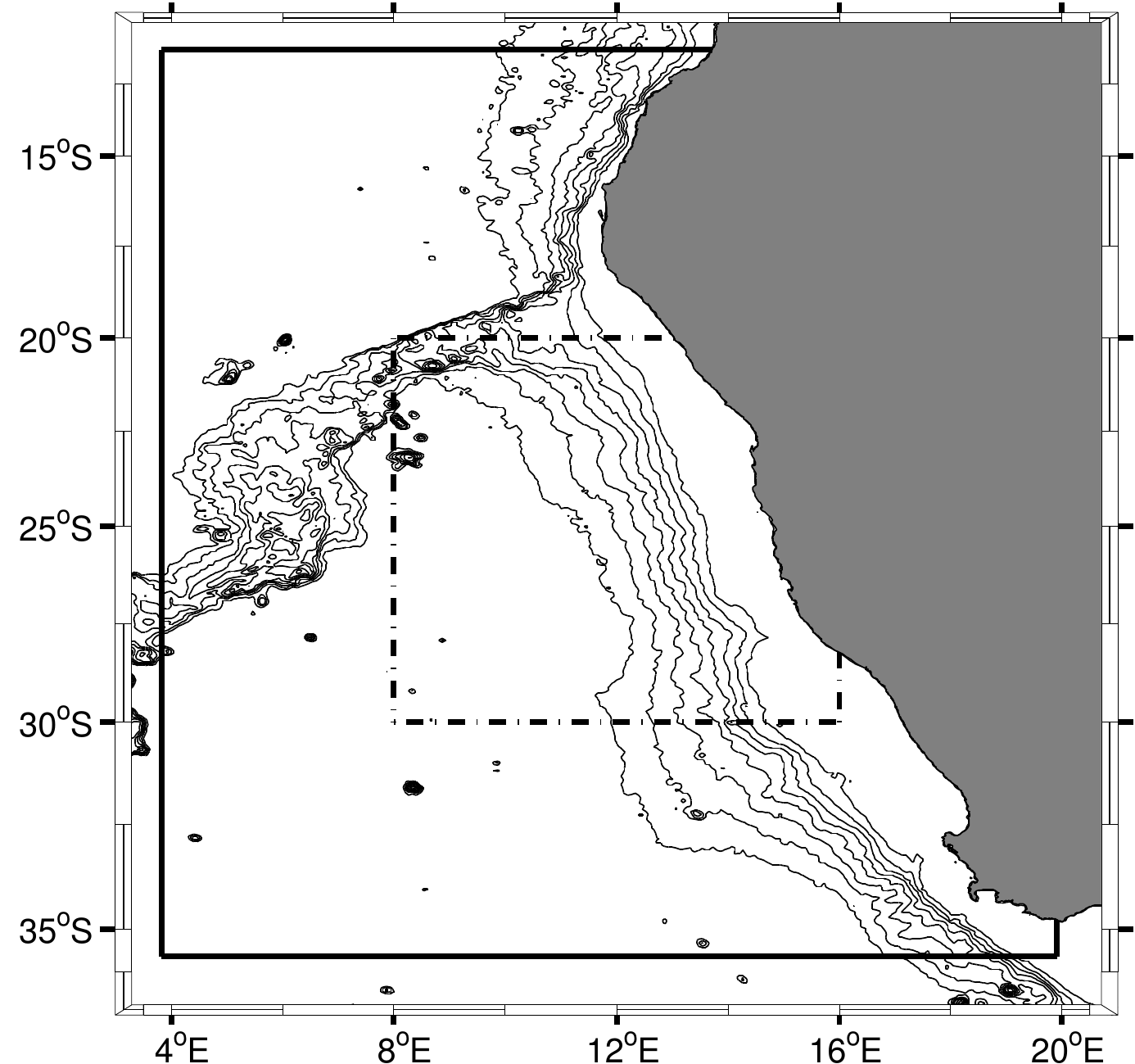}
\caption{
Benguela ocean region. The velocity field domain is limited by the continuous black line.
The FSLE calculation area is limited by the dash-dot black line. Bathymetric contour lines are
from ETOPO1 global relief model \citep{AmanteETOPO1} starting a 0 m depth up to 4000 m at 500 m
interval.}
\label{Fig1}
\end{figure}

\subsection{Finite-Size Lyapunov Exponents.}
\label{subsec:fsle}

In order to study non-asymptotic dispersion processes such as
stretching at finite scales and time intervals, the Finite Size
Lyapunov Exponent  \citep{Aurell1997,Artale1997} is
particularly well suited. It is defined as:
\begin{linenomath*}
 \begin{equation}\label{FSLE_1}
    \lambda=\frac{1}{\tau}\log\frac{\delta_{f}}{\delta_{0}},
 \end{equation}
\end{linenomath*}
where $\tau$ is the time it takes for the separation between
two particles, initially $\delta_{0}$, to reach $\delta_{f}$.
In addition to the dependence on the values of $\delta_0$ and
$\delta_f$, the FSLE depends also on the initial position of
the particles and on the time of deployment. Locations (i.e.
initial positions) leading to high values of this Lyapunov
field identify regions of strong separation between particles,
i.e., regions that will exhibit strong stretching during
evolution, that can be identified with the LCS
\citep{Boffetta2001,dOvidio2004,Joseph2002}.

In principle, for computing FSLEs in three dimensions one just
needs to extend the method of \citet{dOvidio2004}, that is, one
needs to compute the time that fluid particles initially
separated by $\delta_0=[(\delta x_0)^2+ (\delta y_0)^2+(\delta
z_0)^2]^{1/2}$ need to reach a final distance of
$\delta_f=[(\delta x_f)^2+ (\delta y_f)^2+(\delta
z_f)^2]^{1/2}$. The main difficulty in doing this is that in
the ocean vertical displacements (even in upwelling regions)
are much smaller than the horizontal ones, and so do not
contribute significantly to total particle dispersion
\citep{Ozgokmen2011}. By the time the horizontal particle
dispersion has scales of tenths or hundreds of kilometers
(typical mesoscale structures are studied using $\delta_f
\approx 100 km$ \citep{dOvidio2004}), particle dispersion in
the vertical can have at most scales of hundreds of meters and
usually less. This means that the vertical separation will not
contribute significantly to the accumulated distance between
particles. In addition, since length scales in the horizontal
and vertical differ by several orders of magnitude,  one faces
the impossibility of assigning equal $\delta_0$ to the
horizontal and vertical particle pairs. It should be noted
however that these shortcomings arise from the different
scales of length and time that characterize horizontal and
vertical dispersion processes in the ocean, and so should not
be seem as intrinsic limitations of the method. For non-oceanic
flows a direct generalization of FSLEs is straightforward.

Thus, in this paper we implemented a quasi three-dimensional
computation of FSLEs. That is, we make the computation for
every (2d) ocean layer, but where the particle trajectories
calculation use the full 3d velocity field. I.e.,  at each
level (depth) we set $\delta z_O=0$, and the final distance is
computed without taking the vertical distance between
particles. It is important to note that, since we allow the
particles to evolve in the full 3d velocity field, we take into
account vertical quantities such as vertical velocity shear
that may influence the horizontal separation between particle
pairs.

There are other possible approaches to the issue of different
scales in the vertical and horizontal. One way is to assign anisotropic
initial and final displacements in the FSLE calculation 
(i. e., including a $\delta z_0$ and $\delta z_f$ much smaller than 
the horizontal initial and final separations). A second approach is to use 
different weights for the horizontal and vertical separations in the calculations 
of the distance, perhaps in combination with the first. We have cheked both alternatives and
found that, with reasonable choices of initial and final distances and
distance metrics, the results were equivalent to the quasi-3d computation.
The reason is that actual dispersion is primarily
horizontal as commented above.

More in detail, a grid of initial locations $\textbf{x}_{0}$ in
the longitude/latitude/depth geographical space
$(\phi,\theta,z)$, fixing the spatial resolution of the FSLE
field, is set up at time $t$. The horizontal distance among the
grid points, $\delta_0$, was set to $1/36$ degrees ($ \approx
3$ km), i.e. three times finer resolution than the velocity
field \citep{Ismael2011}, and the vertical resolution (distance
between layers) was set to $20$ m in order to have a good
representation of the vertical variations in the FSLE field.
Particles are released from
each grid point and their three dimensional trajectories
calculated. The distances of each particle with respect to the
ones that were initially neighbors at an horizontal distance
$\delta_0$ are monitored until one of the horizontal
separations reaches a value $\delta_{f}$. By integrating the
three dimensional particle trajectories backward and forward in
time, we obtain the two different types of FSLE maps: the
attracting LCS (for the backward), and the repelling LCS
(forward) \citep{dOvidio2004,Joseph2002}. We obtain in this way
FSLE fields with a horizontal spatial resolution given by
$\delta_0$. The final distance $\delta_f$ was set to $100$ km,
which is, as already mentioned, a typical length scale for
mesoscale studies. The trajectories were integrated for a
maximum of $T=178$ days (approximately six months) using an
integration time step of $6$ hours. When a particle reached the
coast or left the velocity field domain, the
FSLE value at its initial position and initial time was set to
zero. If the interparticle horizontal separation remains
smaller than $\delta_{f}$ during all the integration time, then
the FSLE for that location is also set to zero.

The equations of motion that describe the evolution of particle
trajectories are
\begin{linenomath*}
\begin{eqnarray}
\frac{d\phi}{dt} &=& \frac{1}{R_z}\frac{u(\phi,\theta,z,t)}{cos(\theta)},\label{INTX}\\
\frac{d\theta}{dt} &=& \frac{1}{R_z}v(\phi,\theta,z,t),\label{INTY}\\
\frac{dz}{dt} &=& w(\phi,\theta,z,t),\label{INTZ}
\end{eqnarray}
\end{linenomath*}
where $\phi$ is longitude, $\theta$ is latitude and $z$ is the
depth. $R_z$ is the radial coordinate of the moving particle
$R_z=R-z$, with $R=6371$ km the mean Earth radius. For all
practical purposes, $R_z\approx R$. Particle trajectories are
integrated using a $4^{th}$ order Runge-Kutta method. For the
calculations, one needs the (3d) velocity values at the current
location of the particle. Since the six grid nodes surrounding
the particle do not form a regular cube, direct trilinear
interpolation can not be used. Thus, an isoparametric element
formulation is used to map the nodes of the velocity grid
surrounding the particles position to a regular cube, and an
inverse isoparametric mapping scheme \citep{Yuan1994} is used
to find the coordinates of the interpolation point in the
regular cube coordinate system.

\subsection{Lagrangian Coherent Structures.}
\label{LCS}

In 2d, LCS practically coincide with (finite-time) stable and
unstable manifolds of relevant hyperbolic structures in the
flow \citep{Haller2000,Haller2000b,Joseph2002}. The structure
of these last objects in 3d is generally much more complex than
in 2d \citep{Haller2001,Pouransari2010}, and they can be
locally either lines or surfaces. As commented before, however,
vertical motions in the ocean are slow. Thus, at each fluid
parcel the strongest attracting and repelling directions should
be nearly horizontal. This, combined with the incompressibility
property, implies that the most attracting and repelling
regions (i.e. the LCSs) should appear as almost vertical
surfaces, since the attraction or repulsion should occur
normally to the LCS. As a consequence, the LCSs will have a
``curtain-like" geometry, with deviations from the vertical due
to either the orientation of the most attracting or repelling
direction deviating from the horizontal, or when strong vertical
shear produces variations along the vertical in the most
repelling or attracting regions in the flow. We expect the LCS
sheet-like objects to coincide with the strongest hyperbolic
manifolds when these are two dimensional, and to contain the
strongest hyperbolic lines.

The curtain-like geometry of the LCS was already commented in
\citet{Branicki2010b}, \citet{Branicki2010}, or
\citet{Branicki2011}. In the latter paper it was shown that, in a
3d flow, these structures would appear mostly vertical when the
ratio of vertical shear of the horizontal velocity components
to the average horizontal velocities is small. This ratio also
determines the vertical extension of the structures. In
\citet{Branicki2010}, the argument was used to construct a 3d
picture of hyperbolic structures from the computation in a 2d
slice. In the present paper we confirm the curtain-like
geometry of the LCSs, and show that they are relevant to
organize the fluid flow in this realistic 3d oceanic setting.
This is done in the next section by comparing actual particle
trajectories with the computed LCSs.

Differently than 2d, where LCS can be visually identified as
the maxima of the FSLE field, in 3d the ridges are hidden
within the volume data. Thus, one needs to explicitly compute
and extract them, using the definition of LCSs as the ridges of
the FSLEs. A ridge $L$ is a co-dimension 1 orientable,
differentiable manifold (which means that for a
three-dimensional domain $D$, ridges are surfaces) satisfying
the following conditions \citep{Lekien2007}:
\begin{enumerate}
 \item The field $\lambda$ attains a local extremum at $L$.
 \item The direction perpendicular to the ridge is the
     direction of fastest descent of $\lambda$ at $L$.
\end{enumerate}
Mathematically, the two previous requirements can be expressed
as
\begin{linenomath*}
\begin{eqnarray}\label{RREQ}
\mathbf{n}^\mathrm{T}\nabla\lambda &=&0\label{first},\\
\mathbf{n}^\mathrm{T}\mathbf{H}\mathbf{n}=
\min_{\lVert \mathbf{u} \rVert =1}\mathbf{u}^\mathrm{T}\mathbf{H}\mathbf{u} &<& 0,\label{second}
\end{eqnarray}
\end{linenomath*}
where $\nabla\lambda$ is the gradient of the FSLE field $\lambda$,
$\mathbf{n}$ is the unit normal vector to $L$ and
$\mathbf{H}$ is the Hessian matrix of $\lambda$.

The method used to extract the ridges from the scalar field
$\lambda(\mathbf{x}_0,t)$ is from \citet{Schultz2010}. It uses
an earlier \citep{Eberly1994} definition of ridge in the
context of image analysis, as a generalized local maxima of
scalar fields. For a scalar field $f:\mathbb{R}^n \rightarrow
\mathbb{R}$ with gradient $\mathbf{g}=\nabla f$ and Hessian
$\mathbf{H}$, a \textit{d}-dimensional height ridge is given by
the conditions
\begin{linenomath*}
\begin{equation}\label{RIDG1}
\forall_{d<i\leq n} \quad \mathbf{g}^\mathrm{T}\mathbf{e}_i=0 \ \textrm{and}\  \alpha_i<0,  
\end{equation}
\end{linenomath*}
where $\alpha_i, i \in \lbrace 1, 2, \dotsc, n \rbrace$, are
the eigenvalues of $\mathbf{H}$, ordered such that $\alpha_1
\geq \dotsc \geq \alpha_n$, and $\mathbf{e}_i$ is the
eigenvector of $\mathbf{H}$ associated with $\alpha_i$. For
$n=3$, (\ref{RIDG1}) becomes
\begin{linenomath*}
\begin{equation}\label{RIDG2}
    \mathbf{g}^\mathrm{T}\mathbf{e}_3=0 \ \textrm{and}\   \alpha_3<0.
\end{equation}
\end{linenomath*}
This ridge definition is equivalent to the one given by
(\ref{RREQ}) since the unit normal $\mathbf{n}$ is the
eigenvector (when normalized) associated with the minimum eigenvalue of
$\mathbf{H}$. In other words, in $\mathbb{R}^3$ the
$\mathbf{e}_1, \mathbf{e}_2$ eigenvectors point locally along the ridge
and the $\mathbf{e}_3$ eigenvector is orthogonal to it.

The ridges extracted from the backward FSLE map approximate the
attracting LCS, and the ridges extracted from the forward FSLE
map approximate the repelling LCS. The attracting ones are the
more interesting from a physical point of view
\citep{dOvidio2004,dOvidio2009}, since particles (or any
passive scalar driven by the flow) typically approach them and
spread along them, giving rise to filament formation. In the
extraction process it is necessary to specify a threshold $s$
for the ridge strength $|\alpha_3|$, so that ridge points whose
value of $\alpha_3$ is lower (in absolute value) than $s$ are
discarded from the extraction process. Since the ridges are
constructed by triangulations of the set of extracted ridge
points, the $s$ threshold greatly determines the size and shape
of the extracted ridge, by filtering out regions of the ridge
that have low strength. The reader is referred to
\citet{Schultz2010} for details about the ridge extraction
method. The height ridge definition has been used to extract
LCS from FTLE fields in several works (see, among others,
\citet{Sadlo2007}).

\section{Results}
\label{sec:results}

\subsection{Three dimensional FSLE field}
\label{subsec:3d}

The three dimensional FSLE field was calculated for a $30$ day
period starting September 17, with snapshots taken
every $2$ days. The fields were calculated for an area of the
Benguela ocean region between latitudes 20\textdegree S and
30\textdegree S and longitudes 8\textdegree E to 16\textdegree
E (see figure \ref{Fig1}). The area is bounded at NW by the
Walvis Ridge and the continental slope approximately bisects
the region from NW to SE. The western half of the domain has
abyssal depths of about 4000 m. The calculation domain extended
vertically from $20$ up to $580$ m of depth. Both backward and
forward calculations were made in order to extract the
attracting and repelling LCS.

\begin{figure}
\includegraphics[width=\columnwidth]{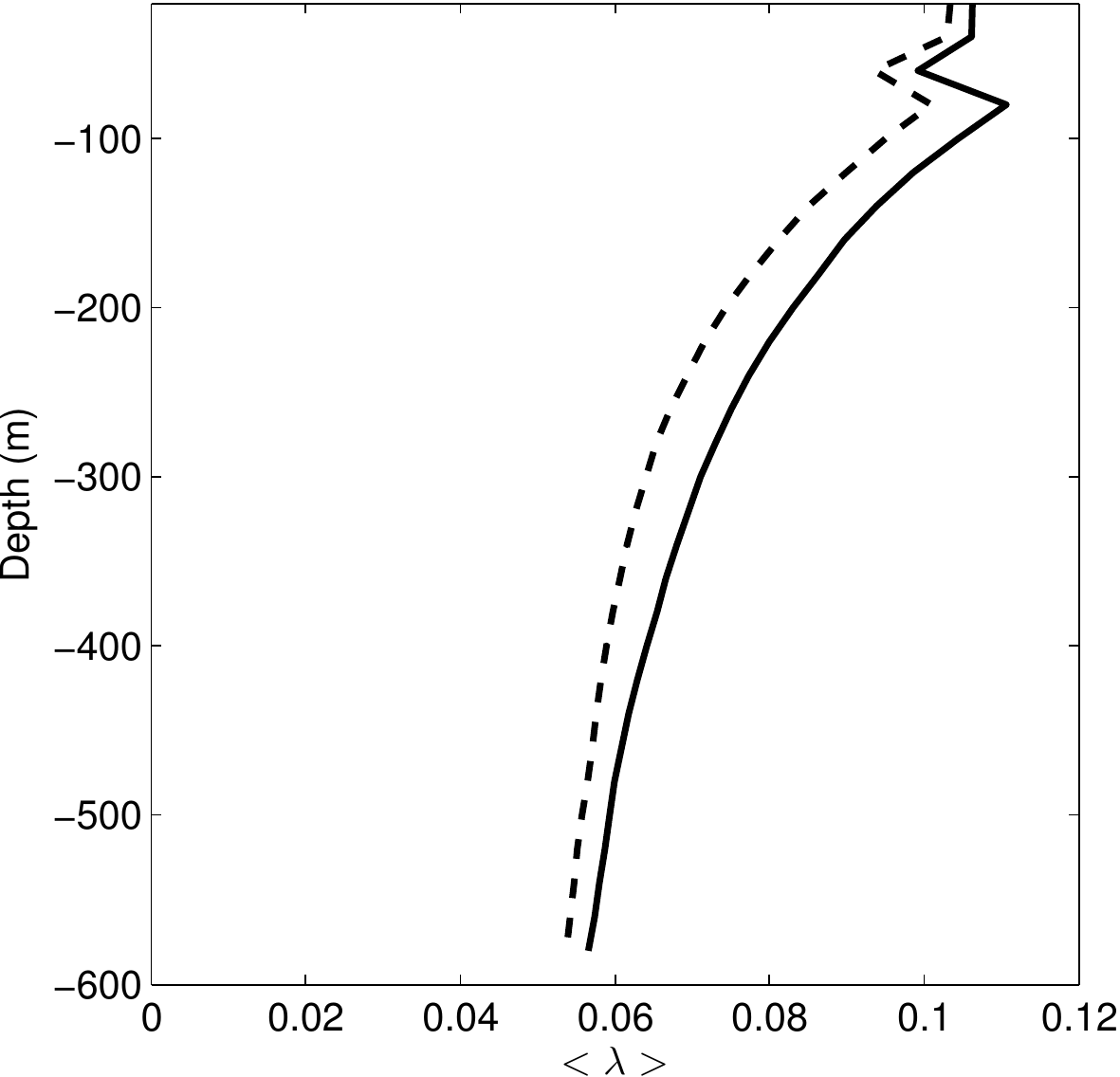}
\caption{
Vertical profile of 30 day average backward
and forward FSLE. The 30 day average field was spatially averaged at each layer over the FSLE calculation area
to produce the vertical profiles. The backward FSLE average is shown in continuous and the forward FSLE is shown in
dashed.}
\label{Fig2}
\end{figure}

Figure \ref{Fig2} displays the vertical profile of the average
FSLE for the 30 day period. There are small differences between
the backward and the forward values due to the different
intervals of time involved in their calculation. But both
profiles have a similar shape and show a general decrease with
depth. There is a notable peak in the profiles at about 100 m
depth that indicates increased mesoscale variability (and
transport, as shown in Sect. \ref{subsec:eddy} at that depth).

\begin{figure}
\includegraphics[width=\columnwidth]{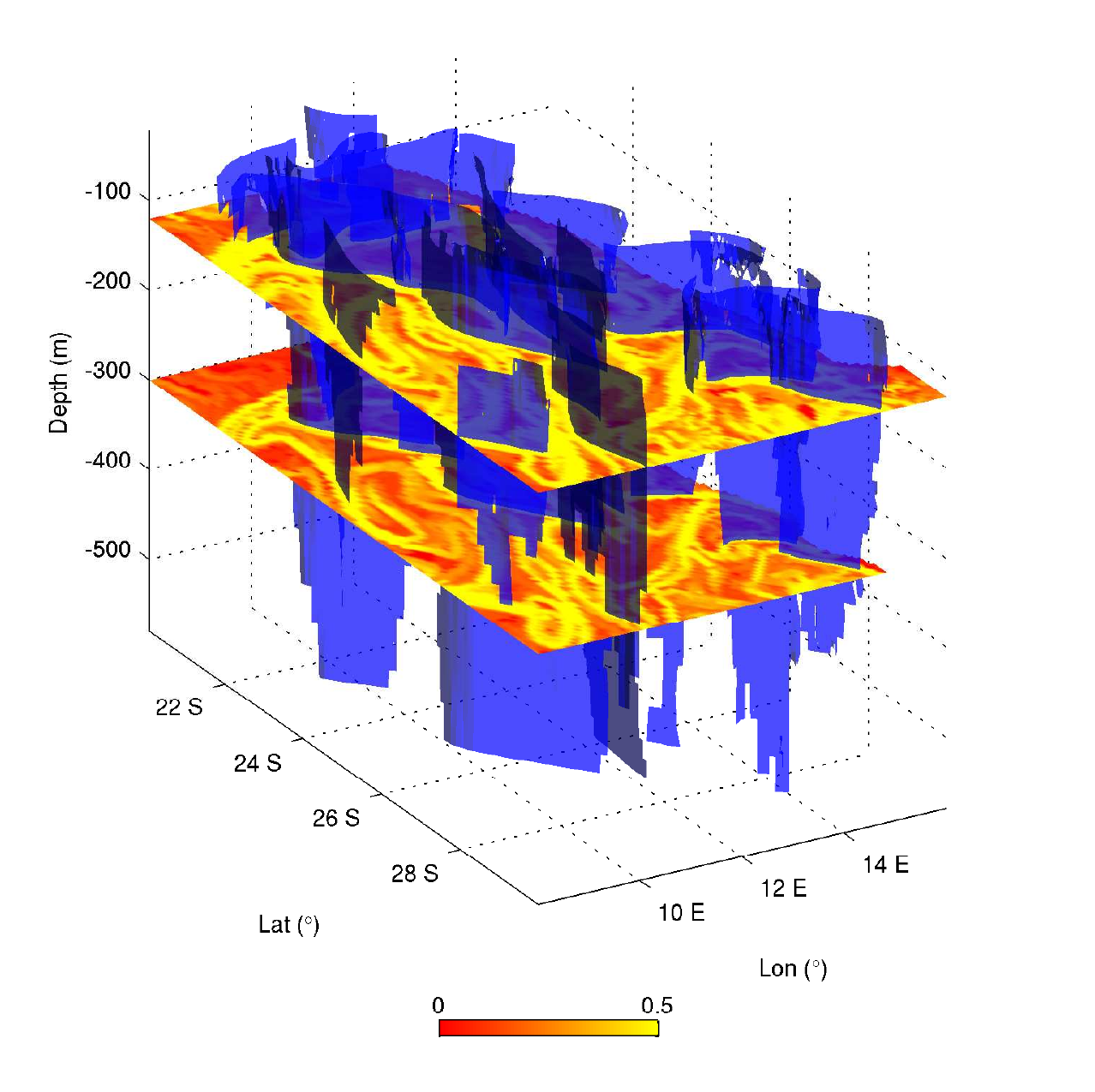}
\caption{
Attracting LCS (blue) for day 1 of the calculation period, together with
horizontal slices of the backward FSLE field at 120 m and 300 m
depth. Colorbar refers to colormap of horizontal slices.
The units of the colorbar are $day^{-1}$.
}
\label{Fig3}
\end{figure}

A snapshot of the attracting LCSs for day 1 of the calculation
period is shown in figure \ref{Fig3}. As expected, the
structures appear as thin vertical curtains, most of them
extending throughout the depth of the calculation domain. The
area is populated with LCS, denoting the intense mesoscale
activity in the Benguela region. As already mentioned, in three
dimensions the ridges are not easily seen, since they are
hidden in the volume data. However the horizontal slices of the
field in figure \ref{Fig3} show that the attracting LCS fall on
the maximum backward FSLE field lines of the 2d slices. The
repelling LCS (not shown) also fall on the maximum forward FSLE
field lines of the 2d slices.

Since the $\lambda$ value of a point on the ridge and the
ridges strength $\alpha_3$ are only related through the
expressions (\ref{RIDG1}) and (\ref{RIDG2}), the relationship
between the two quantities is not direct. This creates a
difficulty in choosing the appropriate strength threshold for
the extraction process. A too small value of $s$ will result in
very small LCS that appear to have little influence on the
dynamics, while a greater value will result in only a partial
rendering of the LCS, limiting the possibility of observing
their real impact on the flow. Computations with several values
of $s$ lead us to the optimum choice $s=20\: day^{-1}m^{-2}$,
meaning that grid nodes with $\alpha_3 < -20\: day^{-1}m^{-2}$ were
filtered out from the LCS triangulation.

We have seen in this section
an example of
how the ridges of the $3d$ FSLE
field, the LCS, distribute in the Benguela ocean region. Their
ubiquity shows their impact on the transport and mixing
properties. In the next section we concentrate on the
properties of a single 3d mesoscale eddy.

\subsection{Study of the dynamics of a relevant mesoscale eddy}
\label{subsec:eddy}

\begin{figure}
\includegraphics[width=\columnwidth]{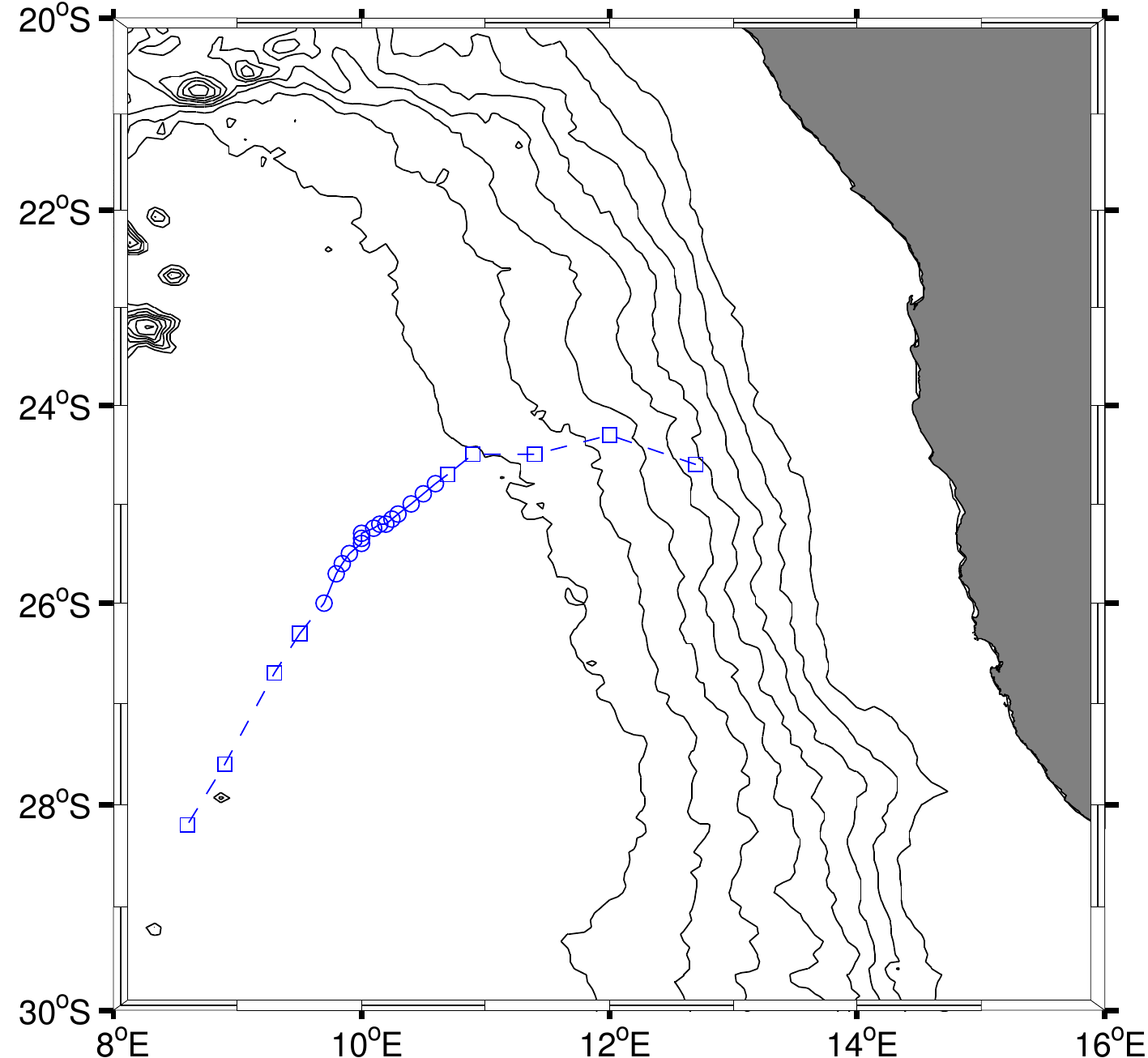}
\caption{
Trajectory (advancing from NE to SW) of the eddy center inside the calculation domain. Circles
indicate the center location during the 30 day FSLE calculation period, and squares previous and
posterior positions. Bathymetric lines same
as in figure \ref{Fig1}.}
\label{Fig4}
\end{figure}

Let us study a prominent cyclonic eddy observed in the data
set. The trajectory of the center of the eddy was tracked and
it is shown in figure \ref{Fig4}. The eddy was apparently
pinched off at the upwelling front. At day $1$ of the FSLE
calculation period its center was located at latitude
24.8\textdegree S and longitude 10.6\textdegree E, leaving the
continental slope, and having a diameter of approximately $100$
km. One may ask: what is its vertical size? is it really a
barrier, at any depth, for particle transport?

To properly answer these questions the eddy, in particular its
frontiers, should be located. From the Eulerian point of view
it is commonly accepted that eddies are delimited by closed
contours of vorticity and that the existence of strong
vorticity gradients prevent the transport in and out of the
eddy. Such transport may occur when the eddy is destroyed or
undergoes strong interactions with other eddies
\citep{Provenzale1999}. In a Lagrangian view point, however, an
eddy can be defined as a region delimited by intersections and
tangencies of LCS, whether in 2d or 3d space. The eddy itself
is an elliptic structure
\citep{Haller2000b,Branicki2010,Branicki2011}. In this
Lagrangian view of an eddy, the transport inhibition to and
from the eddy is now related to the existence of these
transport barriers delimiting the eddy region, which are known
to be quasi impermeable.

\begin{figure*}
  \begin{center}
   \includegraphics[width=0.8\textwidth]{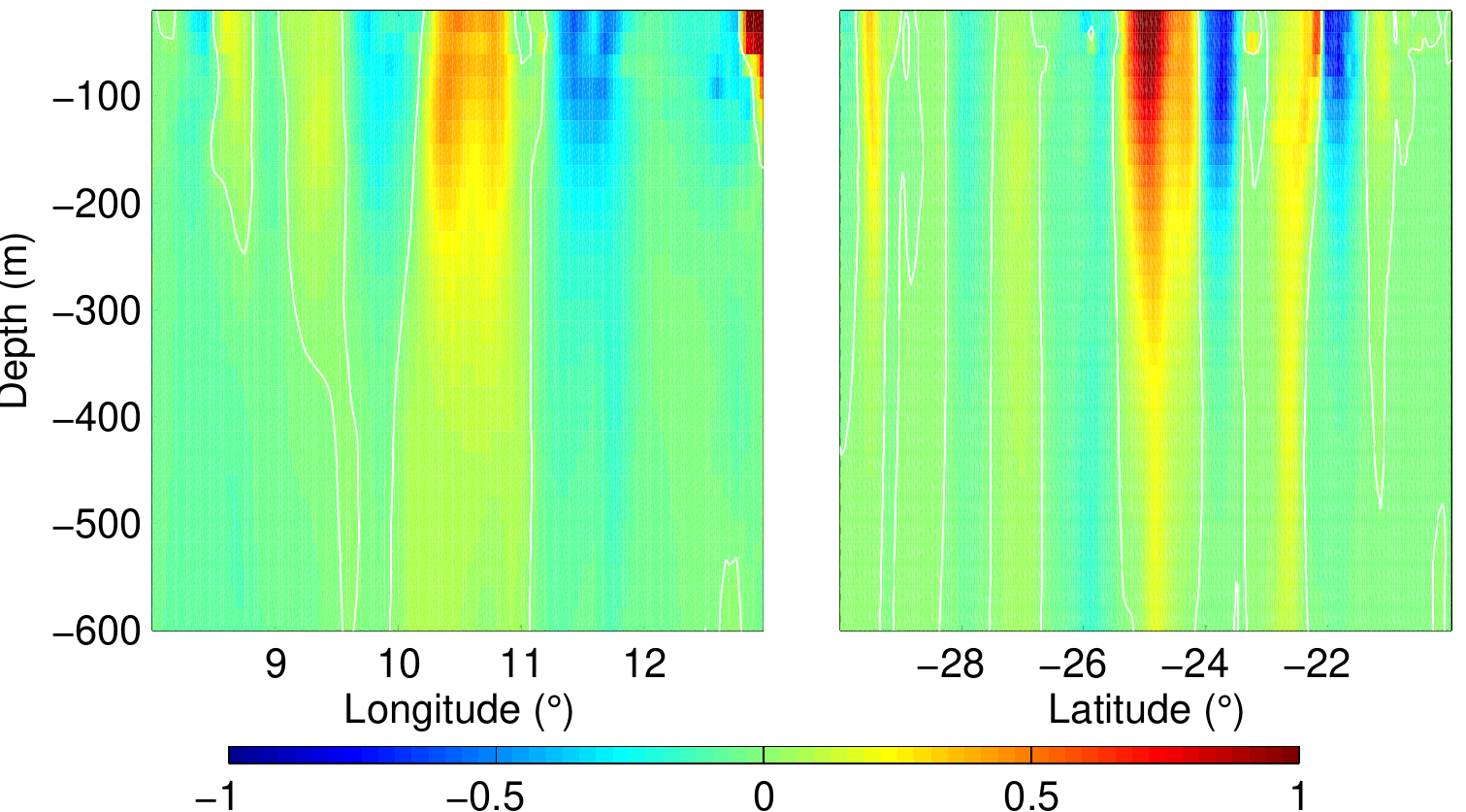}
  \end{center}
  \caption{Colormap of $Q$-criterium. White contours have $Q=0$.
Day $1$ of the $30$ day FSLE calculation period. Left panel:
Latitude $24.5^{\circ}S$; Rigth panel: Longitude $10.5^{\circ}E$. Colorbar
values are $Q\times10^{10}\:s^{-2}$.}
  \label{Fig5}
\end{figure*}

Using the first approach, i.e., the Eulerian view, the vertical
distribution of the $Q$-criteria \citep{Hunt1988,Jeong1995} was
used to determine the vertical extension of the mesoscale eddy.
The $Q$ criterium is a 3d version of the Okubo-Weiss criterium
\citep{Okubo1970,Weiss1991} and measures the relative strength
of vorticity and straining.  In this context, eddies are
defined as regions with positive $Q$, with $Q$ the second
invariant of the velocity gradient tensor
\begin{linenomath*}
\begin{equation}\label{Q1}
    Q=\frac{1}{2}(\|\mathbf{\Omega}\|^2-\|\mathbf{S}\|^2),
\end{equation}
\end{linenomath*}
where
$\|\mathbf{\Omega}\|^2=tr(\mathbf{\Omega}\mathbf{\Omega}^\mathrm{T})$,
$\|\mathbf{S}\|^2=tr(\mathbf{S}\mathbf{S}^\mathrm{T})$ and
$\mathbf{\Omega}$, $\mathbf{S}$ are the antisymmetric and
symmetric components of $\nabla\mathbf{u}$.
Using $Q=0$ as the Eulerian eddy boundary, it can be seen from
Fig. \ref{Fig5} that the eddy extends vertically down to, at
least,  $600$ m.

\begin{figure}
\includegraphics[width=\columnwidth]{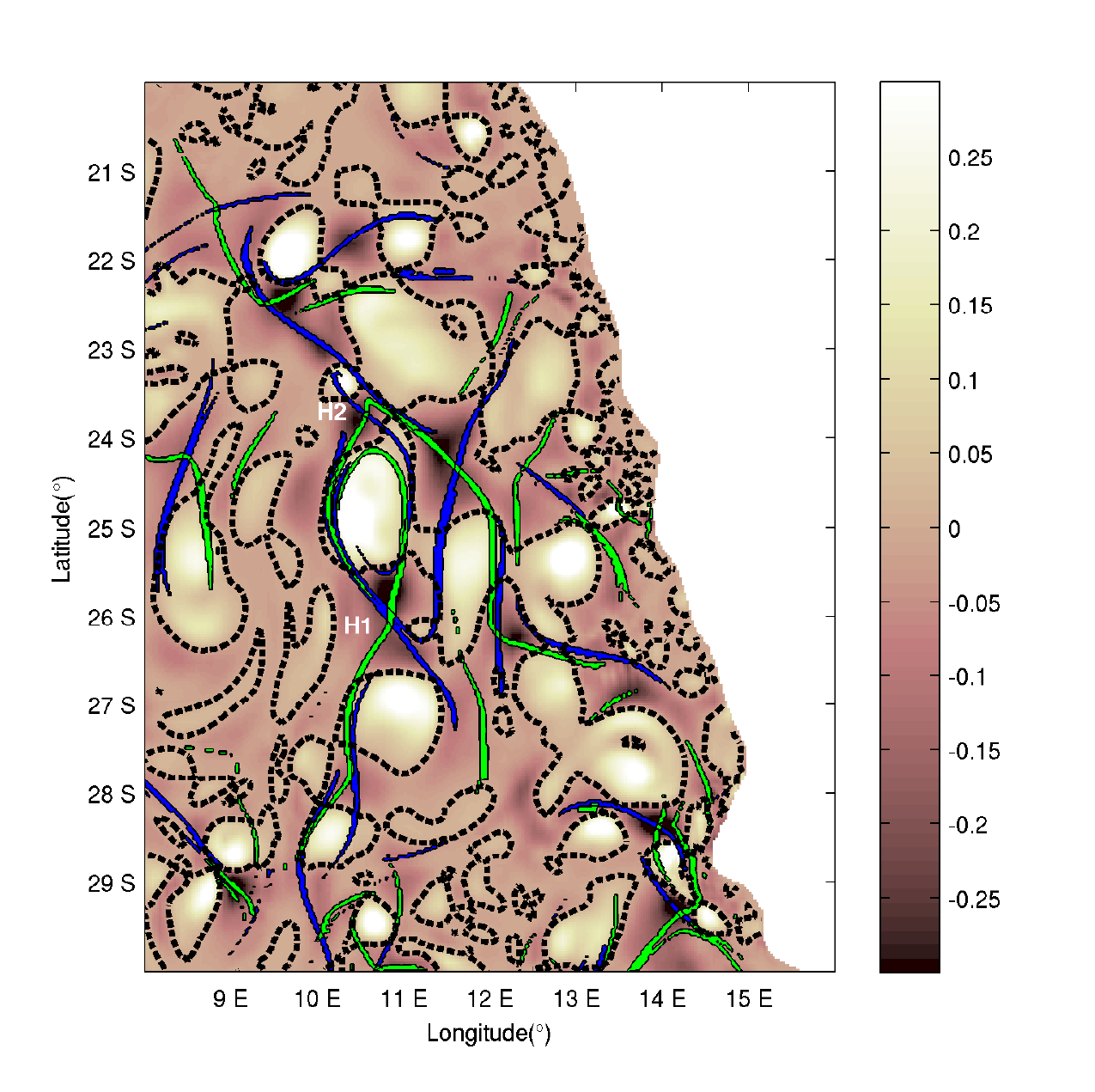}
\caption{$Q$-criterium map at 200 m depth together with patches of backward (blue)
and forward (green) FSLE values. Black dashed lines have $Q=0$. FSLE patches contain the highest
60\% of FSLE values.
Colorbar values are $Q\times10^{10}\:s^{-2}$.The eddy we study
is the clear region in between points H1 and H2.}
\label{Fig6}
\end{figure}

Let us move to the Lagrangian description of eddies, which is
much in the spirit of our study, and will allow us to study
particle transport: eddies can be defined as the {\it region
bounded by intersecting or tangent repelling and attracting
LCS} \citep{Branicki2010,Branicki2011}. Using this criterion,
and first looking at the surface located at 200 m depth, we see
in Fig. \ref{Fig6} that certainly the Eulerian eddy seems to be
located inside the area defined by several intersections and
tangencies of the LCS. This eddy has an approximate diameter of
$100$ km. In the south-north direction there are two
intersections that appear to be hyperbolic points (H1 and H2 in
figure \ref{Fig6}). In the West-East direction, the eddy is
closed by a tangency at the western boundary, and a
intersection of lines at the eastern boundary. The eddy core is
devoid of high FSLE lines, indicating that weak stirring occurs
inside \citep{dOvidio2004}. As additional Eulerian properties,
we note that near or at the intersections H1 and H2 the
$Q$-criterium indicates straining motions. In the case of H2,
figure \ref{Fig5} (right panel) indicates high shear up to 200 m
depth.
The fact that the hyperbolic regions H1 and H2 lie in strain
dominated regions of the flow ($Q<0$) highlights the connection
between hyperbolic particle behavior and instantaneous hyperbolic
regions of the flow. The ridges of the FSLE field, however, do not
remain in the negative $Q$ regions but cross into rotation dominated regions
with $Q>0$. This indicates that there are some differences between the
Eulerian view (Q) and the Lagrangian view (FSLE). It is the latter that
can be understood in terms of particle behaviour as limiting regions
of initial conditions (particles) that stay away from hyperbolic
regions for long enough time \citep{Haller2000b}.

\begin{figure}
  \begin{center}
  \includegraphics[width=\columnwidth]{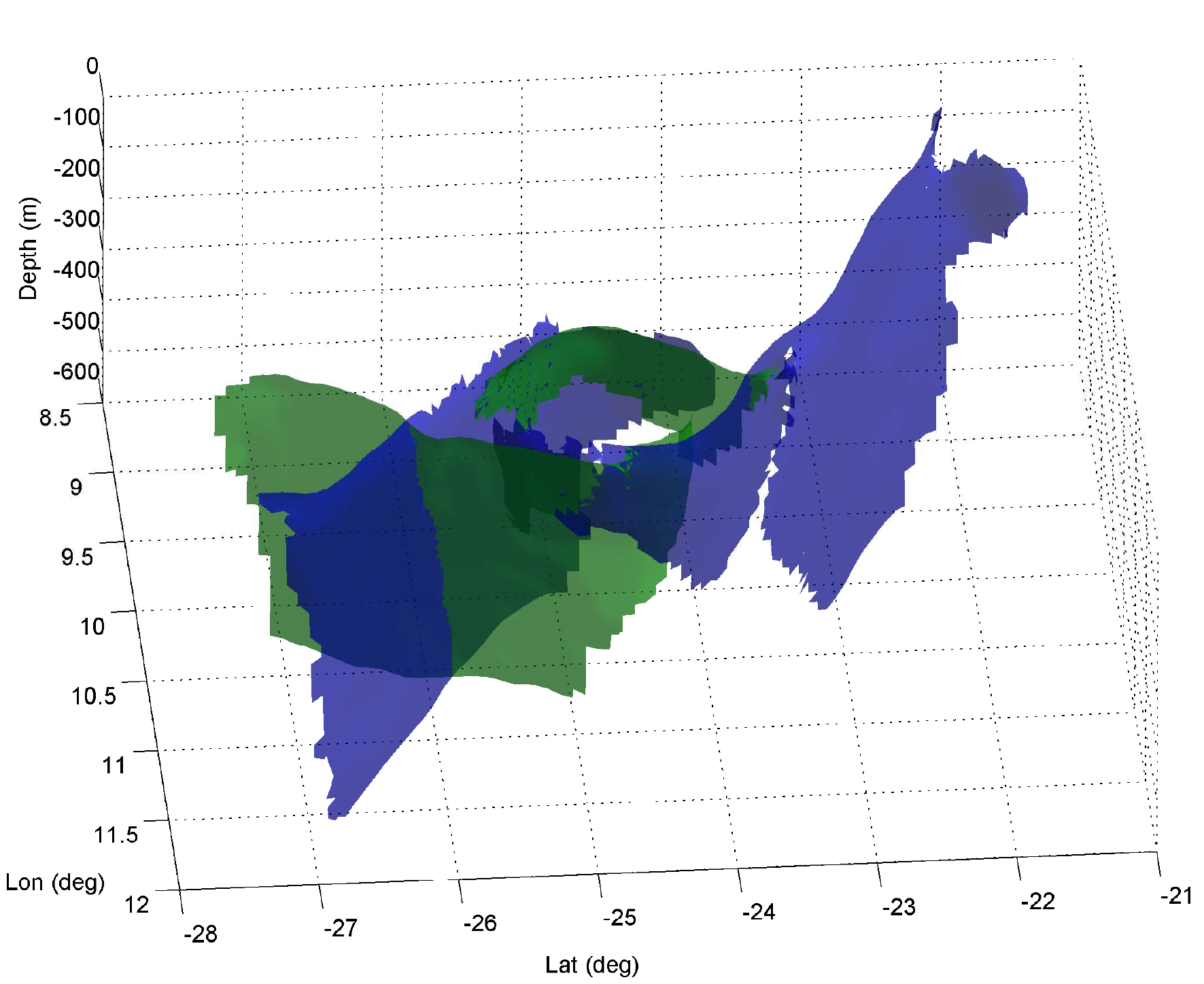}
  \end{center}
  \caption{3d LCSs around the mesoscale eddy at day $1$ of the $30$ day FSLE calculation period.
 Green: repelling LCS; Blue: attracting LCS.}
  \label{Fig7}
\end{figure}

In 3d, the eddy is also surrounded by a set of attracting and
repelling LCS (figure \ref{Fig7}), calculated as explained in
Subsection \ref{LCS}. The lines identified in figure \ref{Fig6}
are now seen to belong to the vertical of these surfaces.

\begin{figure*}[t]
\begin{center}
    \includegraphics[width=0.7\textwidth]{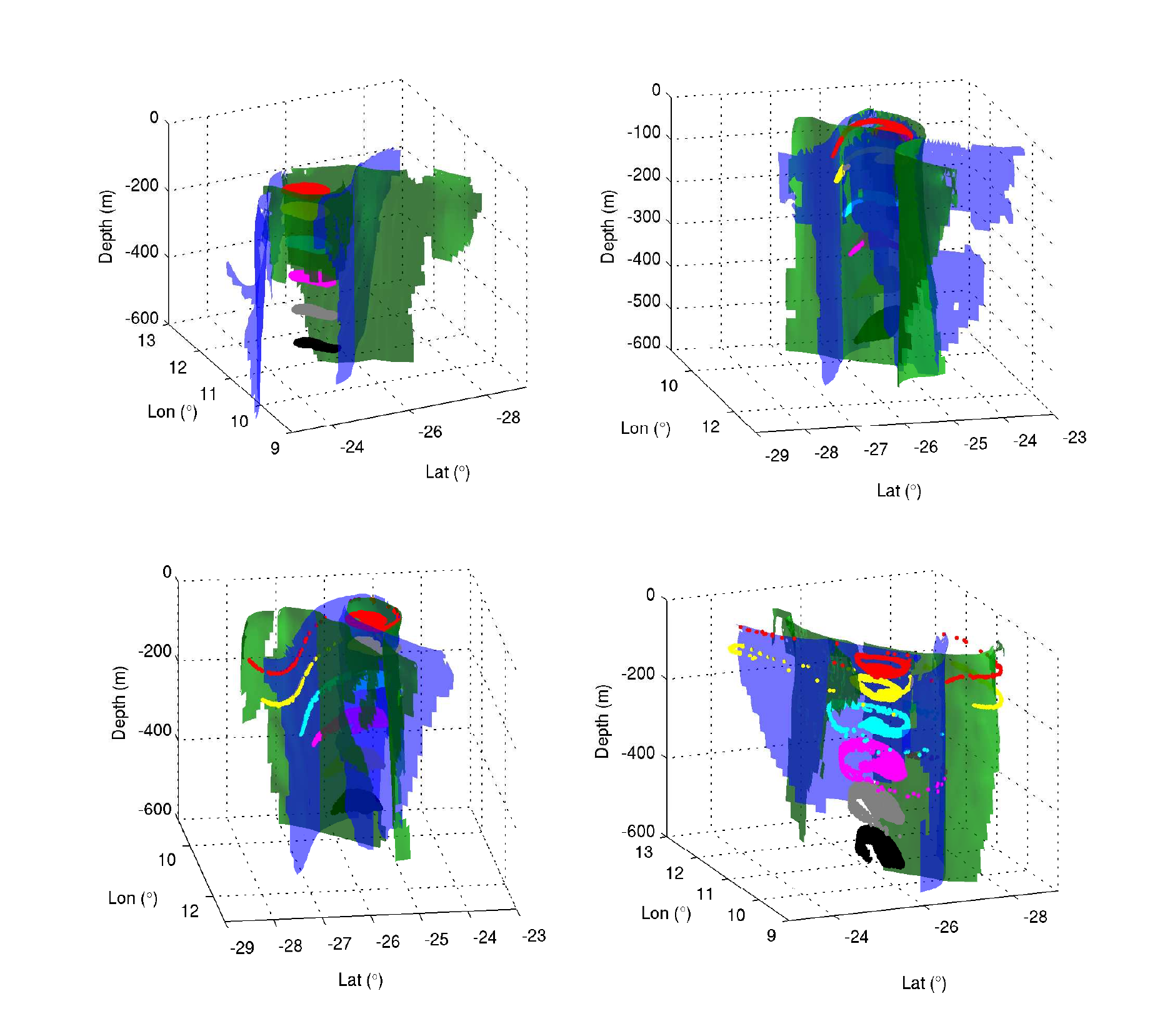}
\end{center}
\caption{Three dimensional view of the evolution of elliptic
patches released at different depths inside of the eddy at day
1 of the 30 day FSLE calculaton period. Top left: day 3; Top
right: day 13; Bottom left: day 19: Bottom right: day 29. Red:
40 m; Yellow: 100 m; Cyan: 200 m; Magenta: 300 m; Grey: 400 m;
Black: 500 m. Attracting LCS are shaded in blue while repelling LCS
are shaded in green.} \label{Fig8}
\end{figure*}

Note that the vertical extent of these surfaces is in part
determined by the strength parameter used in the LCS extraction
process, so their true vertical extension is not clear from the
results presented here. On the south, the closure of the
Lagrangian eddy boundary extends down to the maximum depth of
the calculation domain, but moving northward it is seen that
the LCS shorten their depth. Probably this does not mean that
the eddy is shallower in the North, but rather that the LCS are
losing strength (lower $|\alpha_3|$) and portions of it are
filtered out by the extraction process. In any case, it is seen
that as in two-dimensional calculations, the LCS delimiting the
eddy do not perfectly coincide with its Eulerian boundary
\citep{Joseph2002}, and we expect the Lagrangian view to be
more relevant to address transport questions.

In the next paragraphs we analyze the fluid transport across the eddy
boundary. Some previous results for Lagrangian eddies were
obtained by \citet{Branicki2010} and \citet{Branicki2011}.
Applying the methodology of lobe dynamics and the turnstile
mechanism to eddies pinched off from the Loop Current,
\citet{Branicki2010} observed a net fluid entrainment near the
base of the eddy, and net detrainment near the surface, being
fluid transport in and out of the eddy essentially confined to
the boundary region. Let us see what happens in our setting.

\begin{figure*}[t]
\begin{center}
    \includegraphics[width=0.75\textwidth]{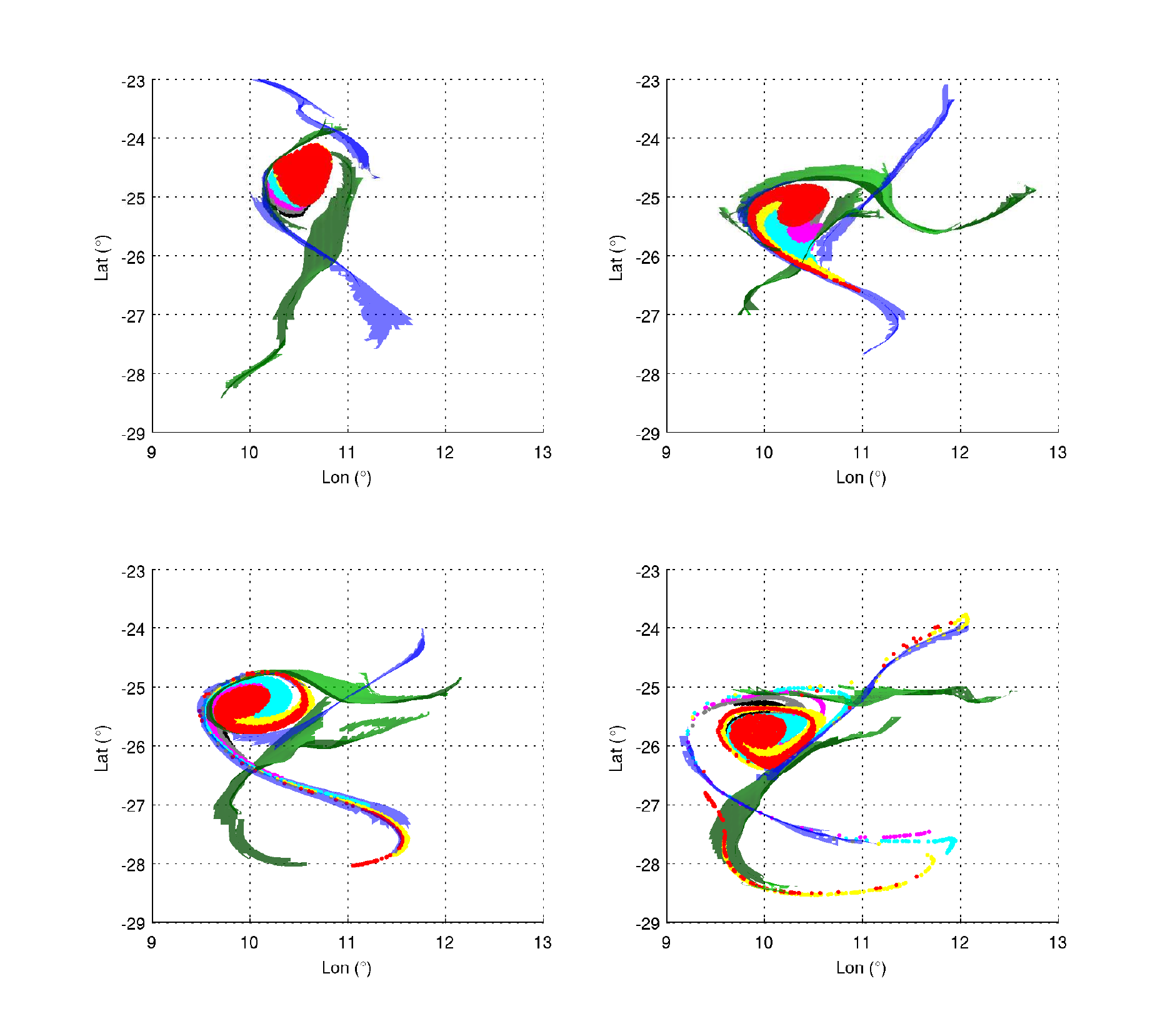}
\end{center}
\caption{Top view of the evolution of particle patches and LCSs
shown in Fig. 8. Top left: day 3; Top right: day 13; Bottom
left: day 19: Bottom right: day 29. Colors as in figure 8.}
\label{Fig9}
\end{figure*}

We consider six sets of $1000$ particles each, that were
released at day $1$ of the FSLE calculation period, and their
trajectories integrated by a fourth-order Runge-Kutta method
with a integration time step of $6$  hours. The sets of
particles were released at depths of $50$, $100$, $200$, $300$,
$400$ and $500$ m. In figure \ref{Fig8} we plot the particle
sets together with the Lagrangian boundaries of the mesoscale
eddy viewed in 3d. A top view is shown in figure \ref{Fig9}. As
expected, vertical displacements are small.

At day $3$ (top left panel of figures \ref{Fig8} and
\ref{Fig9}) it can be seen that there is a differential
rotation (generally cyclonic, i.e. clockwise) between the sets
of particles at different depths. The shallower sets rotate
faster than the deeper ones. This differential rotation of the
fluid particles could be viewed, in a Lagrangian perspective,
as the fact that the attracting and repelling strength of the
LCS that limit the eddy varies with depth. Note that the six
sets of particles are released at the same time and at the same
horizontal positions, and thereby their different behavior is
due to the variations of the LCS properties along depth.

\begin{figure*}[h!]
\begin{center}
    \includegraphics[width=0.75\textwidth]{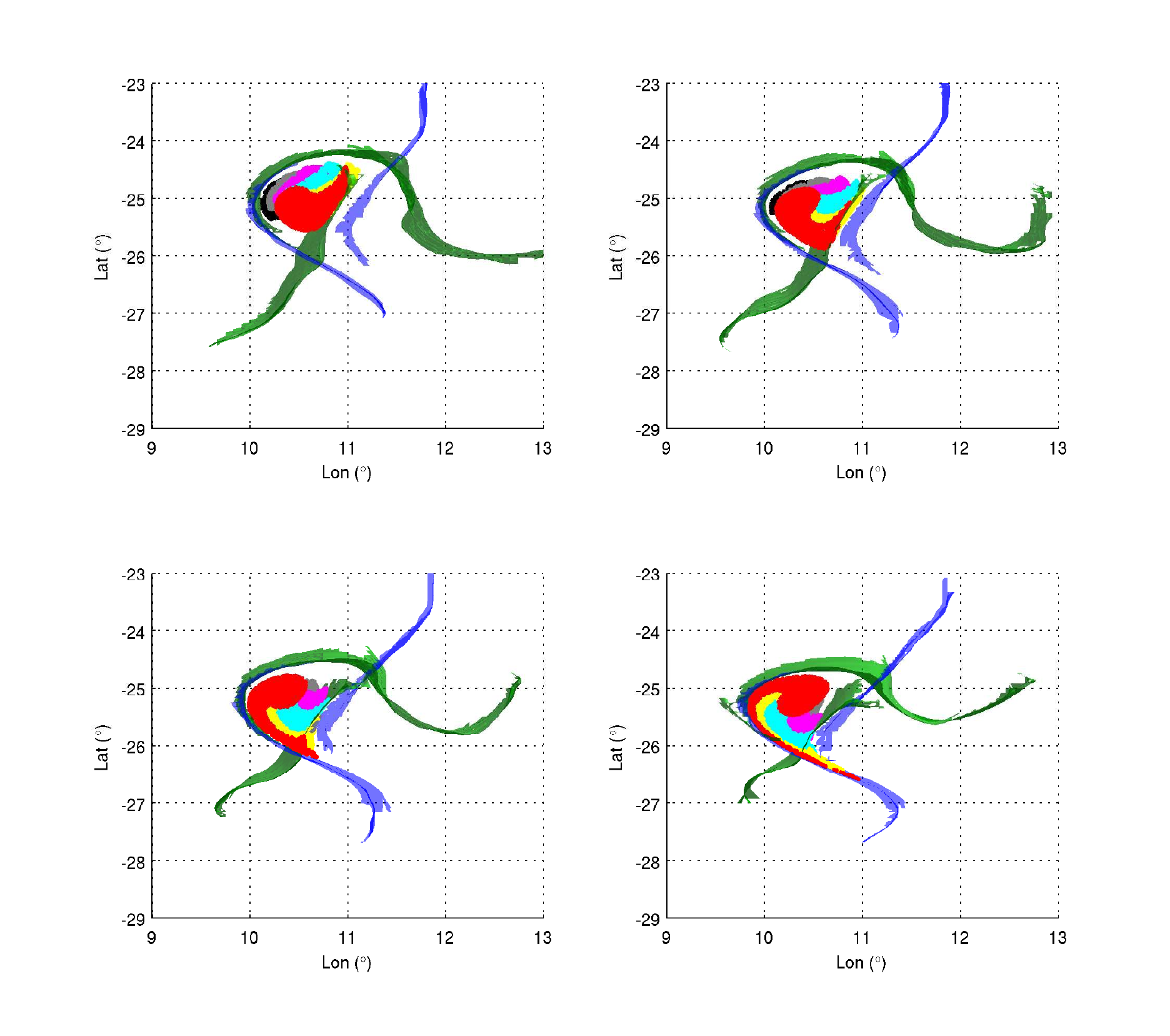}
\end{center}
\caption{Top view of the initial stages of evolution of the
particle patches and LCSs of Figs. 8 and 9. Top left: day 7;
Top right: day 9; Bottom left: day 11: Bottom right: day 13.
Colors as in figure 8.} \label{Fig10}
\end{figure*}

At day $13$ the vortex starts to expel material trough
filamentation (Figs.\ref{Fig8} and \ref{Fig9}, top right
panels). A fraction of the particles approach the southern
boundaries of the eddy from the northeast. Those to the west of
the repelling LCS (green) turn west and recirculate inside the
eddy along the southern attracting LCS (blue). Particles to the
east of the repelling LCS turn east and leave the eddy forming
a filament aligned with an attracting (blue) LCS. At longer
times trajectories in the south of the eddy are influenced by
additional structures associated to a different southern eddy.
At day 29 (bottom right panels) the same process is seen to
have occurred in the northern boundary, with a filament of
particles leaving the eddy along the northern attracting (blue)
LCS. The filamentation seems to begin earlier at shallower
waters than at deeper ones since the length of the expelled
filament diminishes with depth. However all of the expelled
filaments follow the same attracting LCS. Figure \ref{Fig10}
shows the stages previous to filamentation in which the LCS
structure, their tangencies and crossings, and the paths of the
particle patches are more clearly seen. Note that the LCS do
not form fully closed structures and the particles escape the
eddy through their openings. The images suggest lobe-dynamics
processes, but much higher precision in the LCS extraction
would be needed to really see such details.

This filamentation event seems to be the only responsible for
transport of material outside of the eddy, since the rest of
the particles remained inside the eddy boundaries. To get a
rough estimate of the amount of matter expelled in the
filamentation process we tracked the percentage of particles
leaving a circle of diameter 200 km centered on the eddy center.
In Fig. \ref{Fig11} the time evolution of this percentage is
shown for the particle sets released at different depths. The
onset of filamentation is clearly visible around days 9-12 as a
sudden increase in the percentage of particles leaving the
eddy. The percentage is maximum for the particles located at
100 m depth and decreases as the depth increases. At 400 and
500 m depth there are no particles leaving the circle. There is
a clear lag between the onset of filamentation between the
different depths: the onset is simultaneous for the 40 m and
100 m depths but occurs later for larger depths.

\section{Discussion.}

\begin{figure}[ht]
\includegraphics[width=84mm]{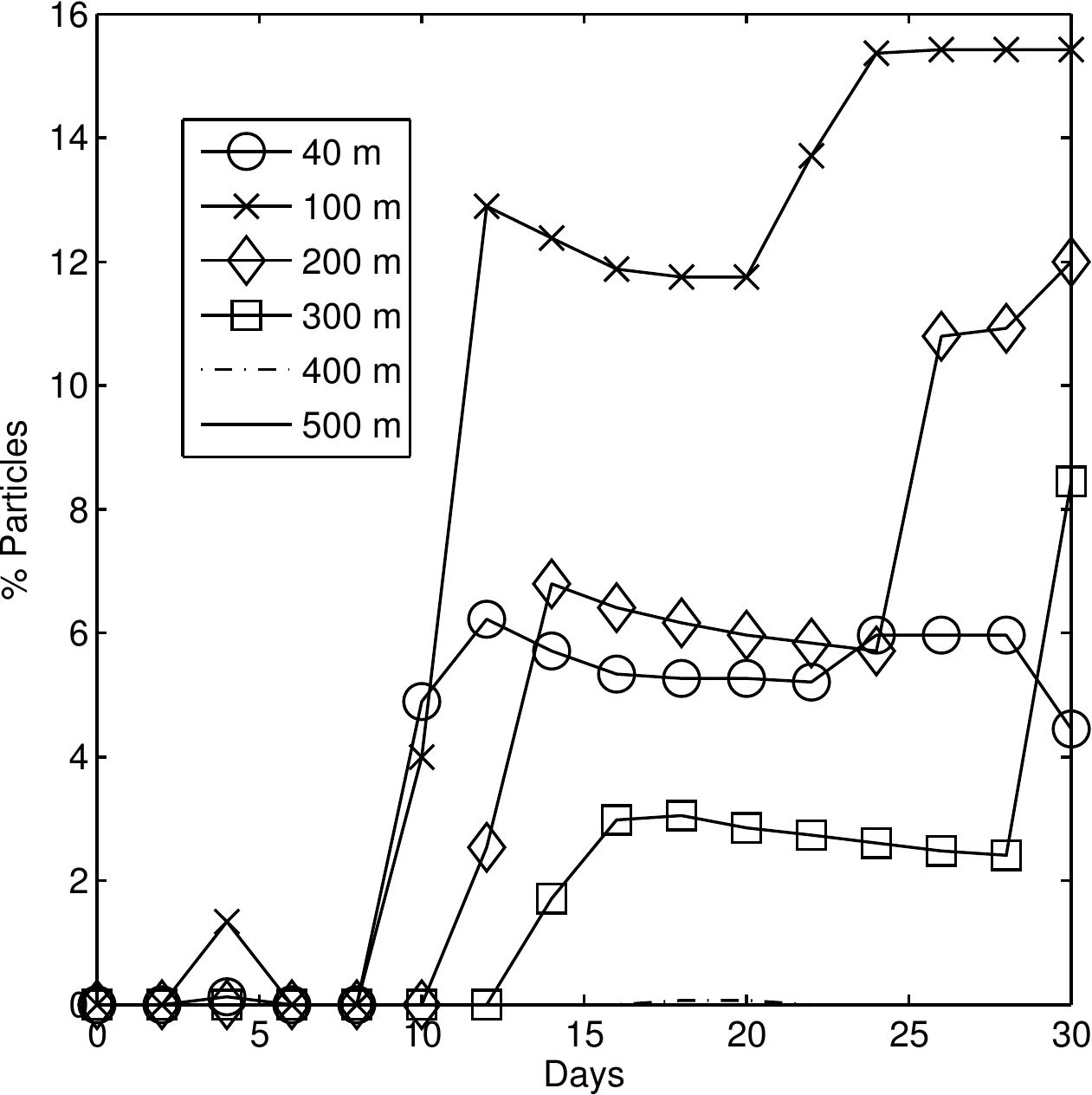}
\caption{Percentage of particles outside a $200$ km diameter
circle centered at the eddy center, as a function of time.
}
\label{Fig11}
\end{figure}

The spatial average of FSLEs defines a measure of stirring and
thus of horizontal mixing between the scales used for its
computation. The larger the average, the larger the mixing
activity \citep{dOvidio2004}. The general trend in the vertical
profiles of the average FSLE (Fig. \ref{Fig3}) shows a
reduction of mesoscale mixing with depth. There is however a
rather interesting peak in this average profile occurring at
100 m, i.e. close to the thermocline. It could be related to
submesoscale processes that occur alongside the mesoscale ones.
Submesoscale is associated to filamentation (the thickness of
filaments is of the order of $10$ km or less), and we have seen
that the filamentation and the associated transport intensity
(Fig. \ref{Fig11}) is higher at $100$ m depth. It is not
clear at the moment what is the precise mechanism responsible
for this increased activity at around 100 m depth (perhaps
associated to instabilities in the mixed layer), but we note
that the intensity of shearing motions (see the $Q$ plots in
\ref{Fig5}) is higher in the top 200 meters. Less intense
filamentation could be caused by reduction of shear in depths
larger than these values.

From an Eulerian perspective, it is thought that vortex
filamentation occurs when the potential vorticity (PV) gradient
aligns itself with the compressional axis of the velocity
field, in strain coordinates
(\citet{Louazel2004};\citet{Lapeyre1999}). This alignment is
accompanied by exponential growth of the PV gradient magnitude.
The fact that the filamentation occurs along the attracting LCS
seems to indicate that this exponential growth of the PV
gradient magnitude occurs across the attracting LCS.

In the specific spatiotemporal area we have studied, and in
particular, for the eddy on which we focussed our analysis, we
have confirmed that the structure of the LCSs is
``curtain-like", so that the strongest attracting and repelling
structures are quasivertical surfaces. Their vertical extension
would depend of the physical transport properties, but it is
also altered by the particular threshold parameter selected to
extract the LCSs. These observations imply that transport and
stirring occurs mainly on the horizontal, which is a reasonable
result considering the disparity between horizontal and
vertical velocities in the ocean, and its stratification.
However, we should mention that our results are not fully
generalizable to all ocean situations, and that any ocean area
or oceanic event should be studied in particular to reveal the
shape of the associated 3d LCS.

Some comments follow about the nature of vertical transport structures.
FSLEs are suited to the identification of hyperbolic structures (structures that exhibit
high rates of transversal stretching or compression in their vicinity).
The question is if one can expect that structures responsible for vertical transport
will also exhibit substantial (vertical) stretching.
This is not so clear in the ocean for the reasons already indicated.
If one considers the case (relevant to our work) of purely isopycnal flow,
then strong vertical stretching would be associated with a rapid divergence of isopycnic surfaces.
In the case of coastal upwelling, for instance, the lifted isopycnic surfaces move vertically
in a coherent fashion, so one should not expect strong vertical divergence of particles flowing
along neighbouring isopycnic surfaces. This is just an example of the fact
that it is possible that coherent vertical motions do not
imply the presence of hyperbolic coherent structures such as those the FSLE may indicate.

Another possible limitation worth mentioning is the velocity 
field resolution and its
relation to the intensity of the vertical velocity. It is accepted that in fronts
or in the eddy periphery, vertical velocities are significantly greater than, 
for instance, in the eddy interior. These zones of enhanced vertical transport 
correspond to submesoscale features that were not adequately captured in 
the velocity field used in this work due to its coarse resolution, since 
submesoscale studies usually have resolutions $<10$ km 
(the literature on this subject is quite
large, so we refer the reader to \cite{Klein2009} and \cite{Levy2008}
).

In any case, a most important point for the LCS we have computed
is that in 3d, as in 2d,
they act as pathways and barriers to transport,
so that they provide a skeleton organizing the transport
processes.

\section{Conclusions}

Three dimensional Lagrangian Coherent Structures were used to
study stirring processes leading to dispersion and mixing at
the mesoscale in the Benguela ocean region. We have computed 3d
Finite Size Lyapunov Exponent fields, and LCSs were identified
with the ridges these fields. LCSs appear as quasivertical
surfaces, so that horizontal cuts of the FSLE fields gives
already a quite accurate vision of the 3d FSLE distribution.
These quasivertical surfaces appear to be coincident
with the maximal lines of the FSLE field (see fig. \ref{Fig3}) so
that surface FSLE maps could be indicative of the position of 3d LCS,
as long as the vertical shear of the velocity 
does not result in a significant deviation
of the LCS with respect to the vertical.
Average FSLE values generally decrease with depth, but we find
a local maximum, and thus enhanced stretching and dispersion,
at about 100 m depth.

We have also analyzed a prominent cyclonic eddy, pinched off
the upwelling front and study the filamentation dynamics in
3d. Lagrangian boundaries of the eddy were made of
intersections and tangencies of attracting and repelling LCS
that apparently emanating from two hyperbolic locations North
and South of the eddy. The LCS are seen to provide pathways and
barriers organizing the transport processes and geometry. This
pattern extends down up to the maximum depth were we calculated
the FSLE fields ($\sim 600$ m), but the exact shape of the
boundary is difficult to determine due to the decrease in ridge
strength with depth. This caused some parts of the LCS not to
be extracted. The inclusion of a variable strength parameter in
the extraction process is an important step to be included in
the future.

The filamentation dynamics, and thus the transport out of the
eddy, showed time lags with increasing depth. This arises from
the vertical variation of the flow field. However the
filamentation occurred along all depths, indicating that in
reality vertical sheets of material are expelled from these
eddies.

Many more additional studies are needed to further clarify the
details of the geometry of the LCSs, their relationships with
finite-time hyperbolic manifolds and three dimensional lobe
dynamics, and specially their interplay with mesoscale and
submesoscale transport and mixing processes.

\section*{Acknowledgements} Financial support from Spanish MICINN
and FEDER through project FISICOS (FIS2007-60327) and from CSIC Intramural
project TurBiD is acknowledged. JHB acknowledges financial support of the
Portuguese FCT (Foundation for Science and Technology) through
the predoctoral grant SFRH/BD/63840/2009. We thank the LEGOS
group for providing us with 3D outputs of the velocity fields
from their coupled BIOBUS/ROMS climatological simulation. The
ridge extraction algorithm of \citet{Schultz2010} is available
in the \texttt{seek} module of the data visualization library
\texttt{Teem} (http://teem.sf.net).

\bibliographystyle{model2-names}


%
%
\end{document}